\documentclass{aa}
\usepackage{txfonts}
\usepackage{graphicx}
\DeclareMathOperator\erf{erf}
\DeclareMathOperator\sgn{sgn}

\usepackage{color}

\begin{document}

\title{Algebraic quantification of an active region contribution to the solar cycle}


\author{Zi-Fan Wang\inst{1,2}
\and Jie Jiang\inst{3,4}
\and Jing-Xiu Wang\inst{2,1}}

\offprints{Jie Jiang, \email{jiejiang@buaa.edu.cn}}

\institute{Key Laboratory of Solar Activity, National Astronomical Observatories, Chinese Academy of Sciences, Beijing 100101, China
\and School of Astronomy and Space Science, University of Chinese Academy of Sciences, Beijing, China
\and School of Space and Environment, Beihang University, Beijing, China
\and Key Laboratory of Space Environment Monitoring and Information Processing of MIIT, Beijing, China}

\abstract
{The solar dipole moment at cycle minimum is considered to be the most reliable precursor with which to determine the amplitude of the subsequent cycle. Numerical simulations of the surface flux transport (SFT) model are widely used to effectively predict the dipole moment at cycle minimum.  An algebraic method was recently proposed to quickly predict the contribution of an active region (AR) to the axial dipole moment at cycle minimum instead of SFT simulations.  The method assumes a bipolar magnetic region (BMR) configuration of ARs, however most ARs are asymmetric in configuration of opposite polarities, or have more complex configurations.  Such ARs evolve significantly differently from those of BMR approximations.}
{We propose a generalized algebraic method to describe the axial dipole contribution of an AR with an arbitrary configuration, and evaluate its effectiveness compared to the BMR-based method. }
{We employ mathematical deductions to obtain the generalized method.  We compare the results of the generalized method with SFT simulations of observed ARs, artificially created BMRs, and ARs with more complex configurations.  We also compare the results with those from the BMR-based method.}
{The generalized method is equivalent to the SFT model, and precisely predicts the contributions of ARs to the dipole moment, but has a much higher computational efficiency.  Although the BMR-based method has similar computational efficiency to the generalized method, it is only accurate for symmetric bipolar ARs.  The BMR-based method systematically overestimates the dipole contributions of asymmetric bipolar ARs, and randomly miscalculates the contributions of more complex ARs.}
{The generalized method provides a quick and precise quantification of the contribution of an
AR to solar cycle evolution, which paves the way for application in physics-based solar cycle predictions.}


\maketitle

\section{Introduction} \label{sec:intro}


Variations of the  long-term magnetic activity of the Sun are the major source of influence on the climate of the Solar System and the surrounding space, affecting the solar wind, the heliosphere, and the related planetary environments \citep{2007AdSpR..40..885M,2007AdSpR..40..891N}.  The long-term development of the solar magnetic field is described by the solar dynamo mechanism, of which the elementary mode is the 22-year magnetic cycle.  As described by the Babcock-Leighton dynamo \citep{1961ApJ...133..572B,1964ApJ...140.1547L,1969ApJ...156....1L}, the initial poloidal field at cycle minimum, which is mostly composed of the polar field, is the source of the toroidal field responsible for the solar active regions during the solar cycle \citep{2013A&A...553A.128J,2015Sci...347.1333C}.  Hence, the polar field at cycle minimum is often regarded as the precursor of the strength of the next cycle \citep{1978GeoRL...5..411S,2017SSRv..210..351W,2018ApJ...863..159J,2020LRSP...17....2P}.  As the polar field is subject to different definitions, the axial dipole moment of the surface magnetic field is often studied as an equivalent.  The dipole moment is introduced by active regions (ARs) that bear initial dipole moments as a result of Joy's law \citep{1919ApJ....49..153H}.  The dipole moment of an
AR evolves as the surface flux transport (SFT) process takes place, and contributes to the dipole moment at cycle minimum, hereafter referred to as the final dipole moment.  Therefore, the contribution of an AR to the final dipole moment quantifies its influence on the long-term development of the solar cycle, as described by the dipole index introduced by \citet{2019ApJ...871...16J}, and by the  active region degree of rogueness (ARDoR) index   introduced by \citet{2020JSWSC..10...46N}.

The contribution of an AR to the solar cycle is commonly obtained by numerical simulations based on the SFT model \citep[e.g.,][]{1985AuJPh..38..999D,1989ApJ...347..529W,1998ApJ...501..866V,2002SoPh..209..287M,2014ApJ...791....5J}.  Meanwhile, it can also be deduced from an algebraic form that is  directly dependent on the properties of the AR, where actual simulations are not necessary.  As shown by the SFT simulations of \citet{2014ApJ...791....5J}, and by the mathematical deductions of \citet{2020JSWSC..10...50P}, the initial dipole introduced by an AR, $D_{i}$, and its contribution to the final dipole, $D_{f}$, follow Eq. (\ref{bmr1})
\begin{equation}\label{bmr1}
  \frac{D_{f}}{D_{i}}=A\exp\left( \frac{-\lambda^{2}}{2\lambda_{R}^{2}} \right),
\end{equation}
where $\lambda$ is the emerging latitude of the AR, and $A$ and $\lambda_{R}$ are constants related to the transport processes.

Equation (\ref{bmr1}) assumes that the ARs emerge in the form of bipolar magnetic regions (BMRs).  Hereafter we refer to Eq. (\ref{bmr1}) as the BMR-based method.  The BMR approximation is unrealistic for ARs.  The two polarities of an AR may be asymmetrical in size and flux, affecting its surface evolution and contribution to the final dipole \citep{2019ApJ...883...24I}.  An AR with more complex configurations, for example a $\delta$-type AR, may deviate from or even be opposite to the predictions of the BMR-based method \citep{2019ApJ...871...16J}.  Generally, ARs with real configurations produce notably different dipole evolutions from those produced by their corresponding BMR approximations \citep{2020SoPh..295..119Y}.  Hence, it is not precise to include the BMR approximation when quantifying the contributions of an AR to the solar cycle, and the configuration of the ARs should be considered.  The current functional form of the BMR-based method cannot include AR configurations directly.  The Gaussian function in Eq. (\ref{bmr1}) assigns the whole AR with a certain latitude.  However, as we decompose a realistic AR into smaller bipolar components, the latitudes of these regions are different.  Hence, an algebraic method is required that includes AR configurations by considering their smaller components.  As the SFT model is linear when nonlinear feedback from magnetic fields to the velocity profiles \citep[e.g., inflows toward activity belts, see][]{2004SoPh..224..217G,2010ApJ...709..301J,2010ApJ...720.1030C,2020JSWSC..10...62N} is not considered, the function describing such elementary components contributing to the final dipole should be linearly addable, so that we can divide a complex AR into its elementary components, evaluate the contributions of these elementary components separately, and add them together.  Hence, such an algebraic method will be capable of predicting the final dipole contribution of an arbitrary AR, and produce a valid quantification of its contribution to the solar cycle without SFT numerical simulations.


We propose a generalized algebraic method to predict the final dipole contribution of ARs with arbitrary magnetic configurations, by generalizing the analytic approach of  \citet{2020JSWSC..10...50P}.  We use SFT simulations of ARs to assess the accuracy of the final dipole contribution predictions of the new generalized method, and the predictions of the previous BMR-based method.  We demonstrate that the generalized method is a theoretical approximation and quick implementation of the SFT simulations, while the previous BMR-based method is only its special case.  As AR configurations are considered, the new generalized method produces better predictions of final dipole than previous methods, and soundly quantifies the contributions of arbitrary ARs to solar cycle evolution.

The present article is organized as follows.  In Sect. \ref{sec:math} we describe the mathematical foundations and deduction of the generalized algebraic method.  In Sect. \ref{sec:sft} we introduce the SFT simulation model which is applied to evaluate the algebraic methods.  In Sect. \ref{sec:evaluation} we evaluate the algebraic methods with the results of SFT simulations.  We discuss and conclude in Sect. \ref{sec:outro}.

\section{Deduction of the algebraic method to predict the dipole contribution of ARs} \label{sec:math}


The basic equation of the SFT model is the radial component of the magnetic induction equation at the solar surface, as shown by Eq. (\ref{SFTeq}), where $B\left(\theta,\phi,t\right)$ is the surface radial field at a certain colatitude $\theta$, longitude $\phi$, and time $t$. Here, $\Omega \left ( \theta  \right )$ denotes the differential rotation, $v\left ( \theta  \right )$  the meridional flow, and $\eta$  the supergranular diffusion.  New flux emergence is inserted as $S\left ( \theta ,\phi ,t \right )$.  As the velocity profiles are usually predetermined and independent of $B$, the equation is linear in terms of $B$, thus allowing the superposition of contributions of different magnetic flux sources.

\begin{equation}\label{SFTeq}
\begin{aligned}
  \frac{\partial B}{\partial t}=-\Omega \left ( \theta  \right )\frac{\partial B}{\partial \phi }-\frac{1}{R_{\odot }\sin \theta }\frac{\partial }{\partial \theta }\left [v\left ( \theta  \right ) B\sin \left ( \theta  \right ) \right ]+\\
  \frac{\eta }{R{_{\odot }}^{2}}\left [\frac{1}{\sin \theta } \frac{\partial }{\partial \theta }\left ( \sin \theta \frac{\partial B}{\partial \theta } \right ) +\frac{1}{\sin ^{2}\theta }\frac{\partial ^{2}B}{\partial \phi^{2}}\right ]+\\
  S\left ( \theta ,\phi ,t \right ).
\end{aligned}
\end{equation}

As an emerged AR evolves according to the SFT processes, the generated cross-equatorial flux is important for the large-scale surface field and the final dipole moment \citep{2013A&A...557A.141C,2017SSRv..210..351W}.  The cross-equatorial flux is a result of the competition between diffusion and meridional flow, and leads to a total flux difference between the two hemispheres.  The flux difference determines the final dipole moment.  The analytic approach of deducing the BMR-based algebraic method by \citet{2020JSWSC..10...50P} follows this concept.  These latter authors discussed the cross-equatorial flux ratio generated by a unipolar flux patch with a Gaussian profile,
\begin{equation}\label{bmr}
B\left(\lambda\right)=\frac{a}{\sigma_{0}}\exp\left[ -\frac{\left( \lambda-\lambda_{0} \right)^{2}}{2\sigma_{0}^{2}} \right],
\end{equation}
at a certain latitude $\lambda_{0}$ with a certain width $\sigma_{0}$.  Assuming a near-equator Cartesian approximation where the first-order Taylor expansion of the meridional flow was considered, \citet{2020JSWSC..10...50P} proved that the fraction of cross-equatorial flux of the Gaussian flux patch on the northern hemisphere is as follows:
\begin{equation}\label{math1}
  f_{crosseq}=\frac{1}{2}\left[1-\erf\left(\lambda_{0}/\sqrt{2}\lambda_{R}\right)\right].
\end{equation}
The parameter $\lambda_{R}$ is
\begin{equation}\label{math2}
  \lambda_{R} = \sqrt{\sigma_{0}^{2}+\frac{\eta}{R^{2}\Delta_{u}}},
\end{equation}
where $R$ is the solar radius, and $\Delta_{u}$ is the derivative of the meridional flow at the equator.  Hence, the cross-equatorial flux is dependent on the initial configuration as well as the transport terms.  They then assumed two flux patches of different polarities, that is, BMRs, and proved Eq. (\ref{bmr1}).

We consider an arbitrary flux region rather than the BMR form.  Because the SFT model shown by Eq. (\ref{SFTeq}) is linear, the contribution of an arbitrary AR to the final dipole moment is a linear addition of the contribution of its smaller parts.  Considering the limit, an arbitrary AR can be decomposed into infinite sizeless ``points''.  Such points are unipolar regions corresponding to magnetogram pixels in reality.  We know that
\begin{equation}\label{math3}
  \lim_{\sigma_{0} \to 0}\left\{ \frac{a}{\sigma_{0}}\exp\left[ -\frac{\left( \lambda-\lambda_{0} \right)^{2}}{2\sigma_{0}^{2}} \right] \right\} = a \sqrt{2\pi}\delta\left( \lambda-\lambda_{0} \right),
\end{equation}
where $\delta\left( \lambda-\lambda_{0} \right)$ is the Dirac $\delta$ function and $a$ is a constant.  From this perspective, Eqs. (\ref{math1}) and (\ref{math2}) also hold for such a ``point''.  The difference is that $\lambda_{R}$ in this case is dependent on transport terms only.

The flux difference between the two hemispheres generated by the point can be deduced from the cross-equatorial flux.  Assuming that at cycle minimum the polar field follows a certain profile determined by the balance of diffusion and meridional flow, the final dipole moment generated by the point $D_{f,point}$ is proportional to the flux difference between the two hemispheres,
\begin{equation}\label{math4}
  D_{f,point}\propto\erf\left(\left|\lambda_{0}\right|/\sqrt{2}\lambda_{R}\right)\sgn\left(\lambda_{0}\right)
.\end{equation}
Here we add a sign function $\sgn\left(\lambda_{0}\right)$ to put different hemispheres into consideration.

The contribution of the whole AR to the final dipole is an integration over the AR,
\begin{equation}\label{math5}
  D_{f}=A_{0} \int\int B\left(\theta,\phi\right)\erf\left(\left|\lambda_{0}\right|/\sqrt{2}\lambda_{R}\right)\sgn\left(\lambda_{0}\right) \sin\theta d\theta d\phi
.\end{equation}
This is the generalized algebraic method we introduce.  Any arbitrary AR can be calculated by integrating over it, and the parameter $\lambda_{R}$, which now equals $\sqrt{\frac{\eta}{R^{2}\Delta_{u}}}$, is independent of AR configuration.  In practice, the numerical integration is carried out following the trapezoidal rule in this article.  In this way, $D_{f}$ is calculated by summing the final dipole contributions of all pixels of the AR.

The proportion ratio $A_{0}$ originates from the proportional relationship between the axial dipole moment and the flux difference between the two hemispheres.  As mentioned above, $A_{0}$ is determined by the balance of diffusion and meridional flow, so it is dependent on the strength of diffusion and the meridional flow profile.  Hence, $A_{0}$ varies with SFT models with different transport parameters, but the proportional relationship is not affected by the exact transport parameters.  $A_{0}$ is a scaling factor when the parameters of the SFT model are given.   We obtain $A_{0}$ by considering the simulations of the SFT model, fitting $A_{0}$ from the results of algebraic methods and SFT simulations.

The generalized algebraic method significantly speeds up the acquisition of final dipole contributions for ARs compared to SFT simulations, because actual numerical calculations are no longer required.  Compared to the BMR-based method \citep{2020JSWSC..10...50P}, the generalized algebraic method requires approximately the same amount of computation time, considering the calculation methods.  The generalized method described by Eq. (\ref{math5}) is an integration over the evaluated AR.  Meanwhile, the BMR-based method requires the initial dipole moment $D_{i}$ of the AR, and so integrating over the AR is also required.  The integration is,
\begin{equation}\label{di}
D_{i}=\frac{3}{4\pi} \int\int B\left(\theta,\phi\right)\cos\theta \sin\theta d\theta d\phi.
\end{equation}
Hence, if the generalized method and the BMR-based method adopt the same numerical integration algorithm, and are applied to the same magnetogram, the computation time will be similar.


\section{Simulations based on the SFT model} \label{sec:sft}

To justify the generalized method, and to compare its effectiveness with the previous BMR-based method, we ran SFT simulations for some ARs, and compared the results of the simulations with those of the prediction methods.  We use an SFT code to solve Eq. (\ref{SFTeq}) based on \citet{2004A&A...426.1075B}.  The spatial resolution of the code is 360$\times$180, and its time interval is 1 day.  The spatial part is decomposed into spherical harmonics up to order 63, and the temporal part is solved with the fourth-order Runge-Kutta method.

The differential rotation profile is adopted from \citet{1983ApJ...270..288S}.  The meridional flow profile is adopted from \citet{1998ApJ...501..866V} with the flow speed set as 11 ms$^{-1}$.  The supergranular diffusion is 500 km$^2$s$^{-1}$.  Under this set of transport parameters, $\lambda_{R}$ for the generalized method is equal to $9^{\circ}.45$.  We use this value in the following evaluations of the generalized method.

Each AR is simulated separately for 10 years  to reach the final state and obtain the final dipole contribution.  To compare with the results of the BMR-based method, each AR is related to its latitude, represented by the average latitude weighted by unsigned flux.  We use two kinds of ARs in the simulations: the ARs from observations of a continuous time period, and the ARs artificially created from predetermined parameters.

For observational ARs, we choose ARs during Carrington rotations (CRs) 2145-2159, as these ARs are recognized as important in the cycle development of cycle 24, and, as shown by the SFT
simulations of \citet{2020ApJ...904...62W}, contribute a large fraction of the final dipole.  Here we adopt the 84 ARs identified and used in the work of these latter authors, in which the characteristics and evolution properties are discussed in detail.  Some of these ARs bear complex configurations instead of simple BMR form, as presented by \citet{2020ApJ...904...62W}.

For artificially created ARs, we consider bipolar ARs and ARs with complex configurations.  The bipolar ARs are composed of two Gaussian flux patches of opposite polarity \citep{2004A&A...426.1075B,2014ApJ...791....5J}.  The maximum magnetic field strength is set to 250 G, and the tilt angles of the bipolar ARs are determined by Joy's law, $\alpha=k\sqrt{\left|\lambda\right|}$, where $\alpha$ is the tilt angle and $k=1.3$ is a typical value of the Joy's coefficient \citep{2010ApJ...719..264C,2020ApJ...900...19J}.  The latitudes considered range from 0$^{\circ}$ to 35$^{\circ}$ with a 5$^{\circ}$ interval, covering eight latitudes in total.  In this form, the bipolar ARs at 0$^{\circ}$ latitude have no tilts, and so their initial dipoles are also zero.  In order to show the influence of AR configurations to the prediction methods, we include morphological asymmetries of the two polarities, which were introduced by \citet{2019ApJ...883...24I}.  With an asymmetry factor $f_{asym}$, the maximum field strengths $B_{max}$ and the widths of the Gaussian profiles $\sigma_{0}$ are modified as,
\begin{equation}\label{asym}
  f_{asym}=\frac{B_{max,L}}{B_{max,F}} = \left(\frac{\sigma_{0,F}}{\sigma_{0,L}}\right)^{2}
,\end{equation}
where the subscript \emph{L} indicates the leading polarity and the subscript \emph{F} indicates the following polarity, identical to the factor defined by \citet{2019ApJ...883...24I}.  In this way, the following polarity is more diffusive.  We consider the cases where $f_{asym}$ equals 1, 2, and 3, and refer to them as the symmetric, weakly asymmetric, and strongly asymmetric cases, respectively.

To compare with the results of bipolar ARs, we also create ARs with more complex configurations at certain latitudes.  The ARs are modeled after NOAA AR 12673, which was studied in detail by \citet{2019ApJ...871...16J}.  As shown by their work, AR 12673 is a typical case where the SFT model produces significantly different results from the BMR-based model as a result of its complex configuration.  Here we flip it into the northern hemisphere to compare with the aforementioned artificially generated bipolar ARs, and shift its latitudinal locations of each pixel in the equal-latitude projection.  The latitudes of the complex ARs, defined as the area-weighted flux centers, are now also from 0$^{\circ}$ to 35$^{\circ}$ with a 5$^{\circ}$ interval.  




\section{Evaluating the generalized algebraic method with SFT simulations} \label{sec:evaluation}

\subsection{Evaluation using observed ARs} \label{subsec:evaars}

We compare the simulated final dipole moment for the observed ARs during CRs 2145-2159 with the results from the algebraic prediction methods.  For the generalized algebraic method, we compute the integration for each isolated AR over the whole solar surface, with the constant $\lambda_{R}$ given in Sect. \ref{sec:sft}.  As the SFT simulations decompose the surface magnetic field into spherical harmonics up to order 63, we also apply the integration over the ARs with their spherical harmonics components up to order 63, while the higher orders are not included in the integration for consistency in numerical methods. The value of the coefficient $A_{0}$ in Eq. (\ref{math5}) is 0.21, which is obtained by linear fitting the simulated dipole contributions to the results of Eq. (\ref{math5}). 

\begin{figure*}
 \centering
\includegraphics[width=0.70\textwidth]{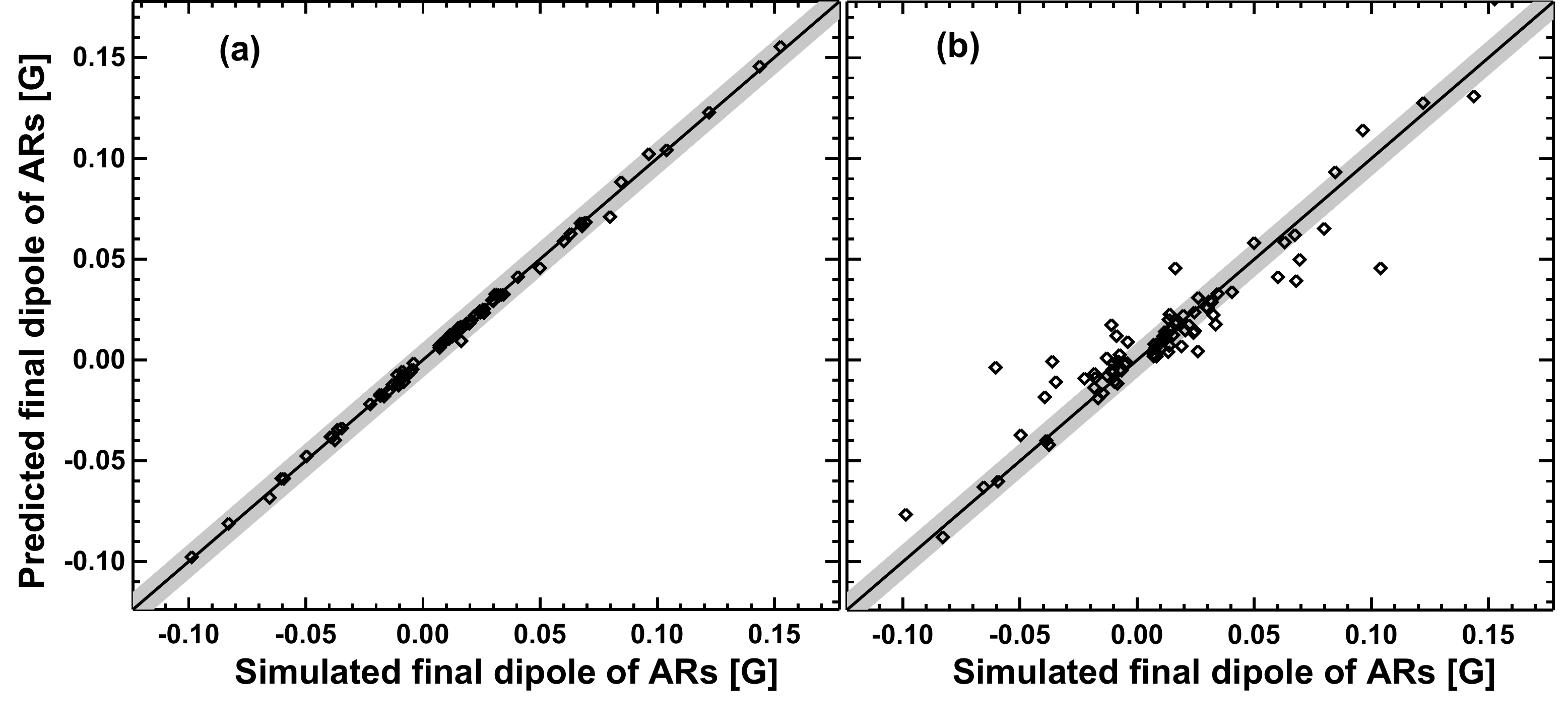}
\caption{Comparison of the  simulated final dipoles of the ARs and those predicted using the algebraic methods.  Each diamond represents an AR during CRs 2145-2159. The horizontal x axis represents the final dipole produced by SFT simulations.  The vertical y axis represents the final dipole predicted by algebraic methods. Panel (a) presents the results of the generalized method that we introduce; panel (b) presents the results of the BMR-based method.  The diagonal line represents $y=x$.  The gray shaded area represents the region where the final dipole produced by the algebraic methods shows no greater than 0.009G deviation from the result produced by SFT simulations.\label{fig:dipar}}
\end{figure*}

The results of the simulated final dipole moment and corresponding final dipole moment predicted by the generalized method for each AR are displayed in Fig. \ref{fig:dipar}(a).  As shown, the predictions of the generalized algebraic method are almost identical to the results of SFT simulations.  The maximum difference between the predicted and simulated final dipole is 0.009G ($11\%$ of the dipole contribution from the AR that generates the maximum difference), and is represented by the gray shaded area.  As this area shows, the error of the final dipole is actually not directly related to the absolute value of the final dipole.  The error may originate from the deviation from the near-equator Cartesian assumption and the lowest order Taylor approximation of the meridional flow when we deduce the algebraic method.  The method is a practical and effective simplification of the SFT model, as expected by the mathematical arguments in Sect. \ref{sec:math}.

We compare the results of the generalized algebraic method with the BMR-based method.  To utilize the BMR-based method, we calculate the initial dipole moment $D_{i}$ for each AR.  Similarly, the integration is calculated based on the ARs with their spherical harmonics components up to order 63.  The initial dipole moments are multiplied by corresponding contribution factor $D_{f}/D_{i}$.  According to Eq. (\ref{math2}), $\lambda_{R}$ depends on $\sigma_{0}$, the initial Gaussian width of the polarities, which is not a well-defined quantity for ARs with real configurations.  Hence, we adopt the value $\lambda_{R}=10^{\circ}.06$ fitted from the symmetric bipolar ARs in Sect. \ref{subsec:evabmrs}, in which case $\lambda_{R}$ follows the definition of Eq. \ref{math2}.  The results of the BMR-based method are displayed in Fig. \ref{fig:dipar}(b).  The scatter of the points is considerably larger than from the generalized method.  The maximum difference between predictions and simulation is now 0.06G ($56\%$ of the dipole contribution of the AR), produced by NOAA AR 11990 on CR 2157.  AR 11990 is located within the corresponding area of AR 11967, which is a complex $\beta\gamma\delta$-type AR on CR 2156.  The remnant of AR 11967 causes the identified AR 11990 to be very asymmetric, resulting in a large deviation.  This shows that the BMR-based method is not accurate when the flux emergence is in more complex forms.

\begin{figure*}
\centering
\includegraphics[width=0.5\textwidth]{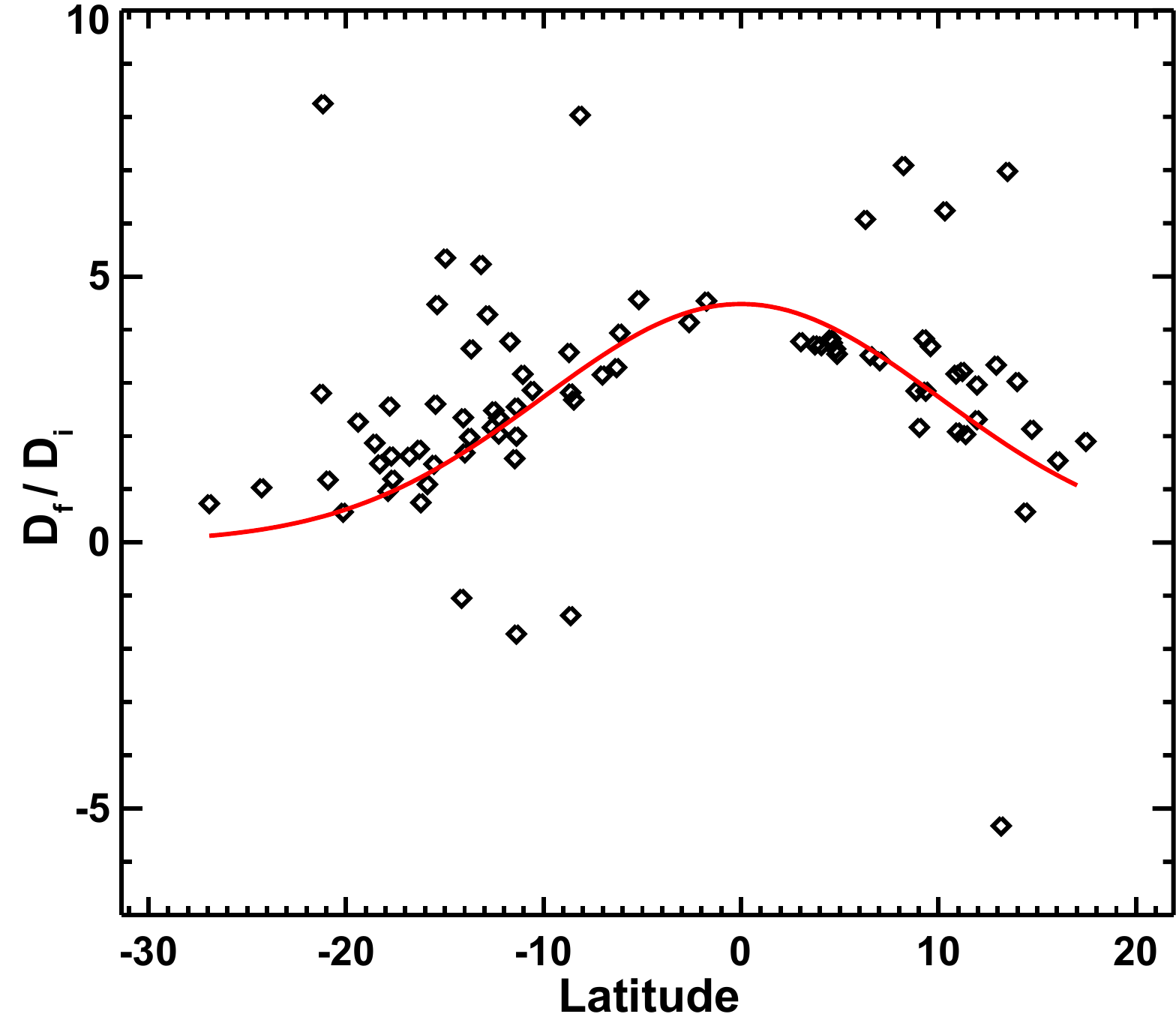}
\caption{Final dipole contribution factor $D_{f}/D_{i}$, with each diamond representing an AR during CRs 2145-2159.  The red trend line representing Eq. (\ref{bmr1}) is plotted by setting $\lambda_{R}=10^{\circ}.06$.\label{fig:fvaluear}}
\end{figure*}

To further demonstrate the deviation of $D_{f}/D_{i}$ with realistic AR configurations from the values with the BMR approximations, we plot $D_{f}/D_{i}$ for each AR in Fig. \ref{fig:fvaluear}.  As shown, while there appears to be a Gaussian trend expressed by Eq. (\ref{bmr1}) and indicated by the red curve, many points deviate from the trend, which is similar to the results of \citet{2020SoPh..295..119Y}.  Hence, the BMR-based method is limited for realistic ARs, while the generalized method produces more reliable predictions as AR configurations are considered.  Please note that although Fig. \ref{fig:dipar}(b) and Fig. \ref{fig:fvaluear} show that the BMR-based method tends to give lower predicted final dipole moments, this is not necessary the case.  The trend results from the limited ARs considered here.  When a substantial number of ARs are considered, there is no trend, as shown by Fig. 8 of \citet{2020SoPh..295..119Y}.  The BMR-based method could give a stronger or lower predicted final dipole moment depending on the configuration of the AR; we discuss this further in Sect. \ref{subsec:evabmrs}.



\subsection{Evaluation using artificially created ARs} \label{subsec:evabmrs}
\begin{figure*}
\centering
\includegraphics[width=0.8\textwidth]{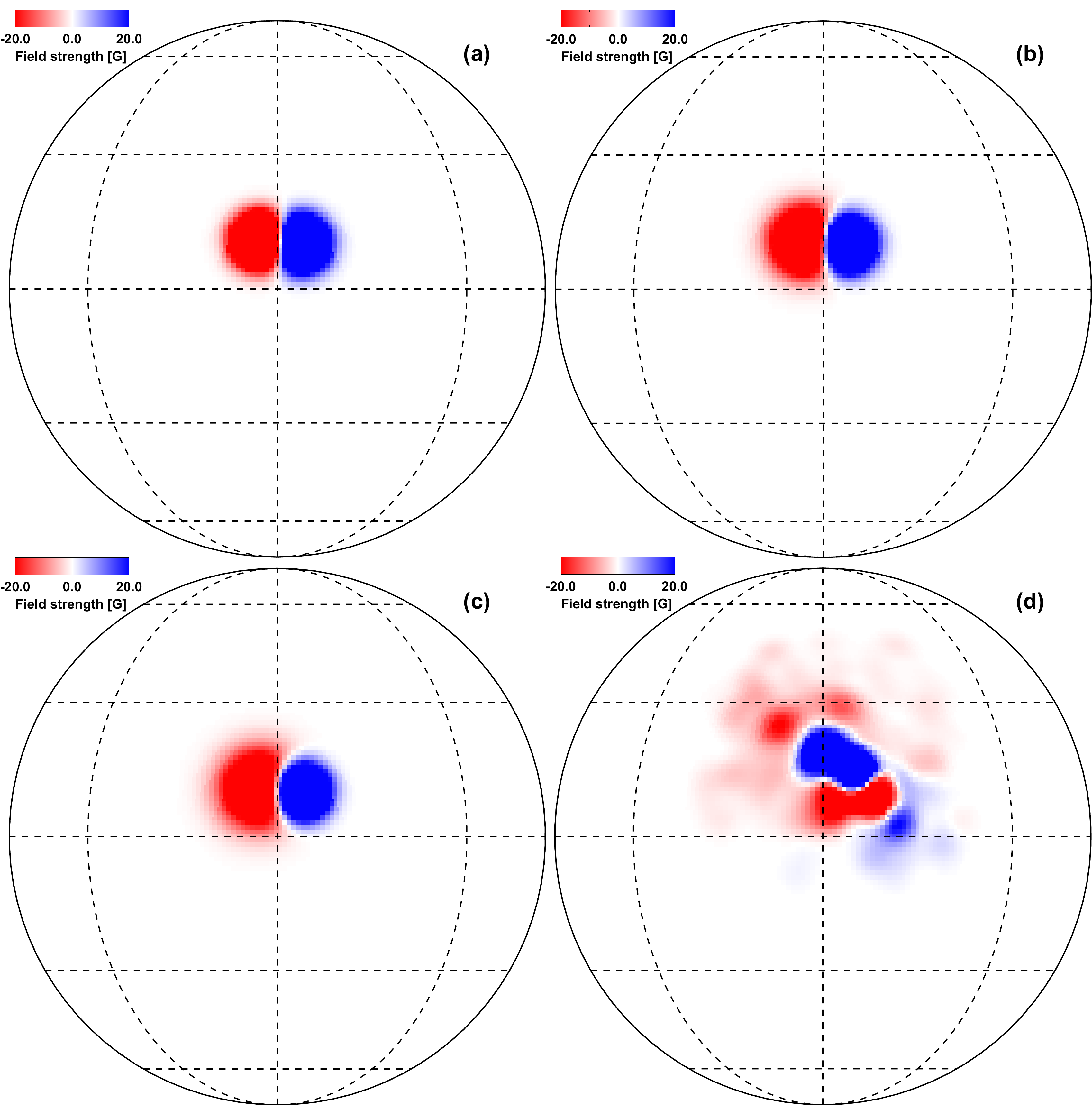}
\caption{Examples of the artificially created ARs with different types of configurations: (a) symmetric bipolar AR; (b) weakly asymmetric bipolar AR; (c) strongly asymmetric bipolar AR; (d) complex AR. The examples are orthographically projected.\label{fig:bmrs}}
\end{figure*}

In order to evaluate the generalized method, and to explicitly show the difference between the generalized method and the previous BMR-based method, we employ both methods to predict the four sets of artificially created ARs.  A set of examples of the ARs is displayed in Fig. \ref{fig:bmrs}.  As shown, the ARs with different degrees of asymmetry differ in their following polarities, while the complex AR is completely different from the bipolar ARs.

\begin{figure*}
\centering
\includegraphics[width=0.5\textwidth]{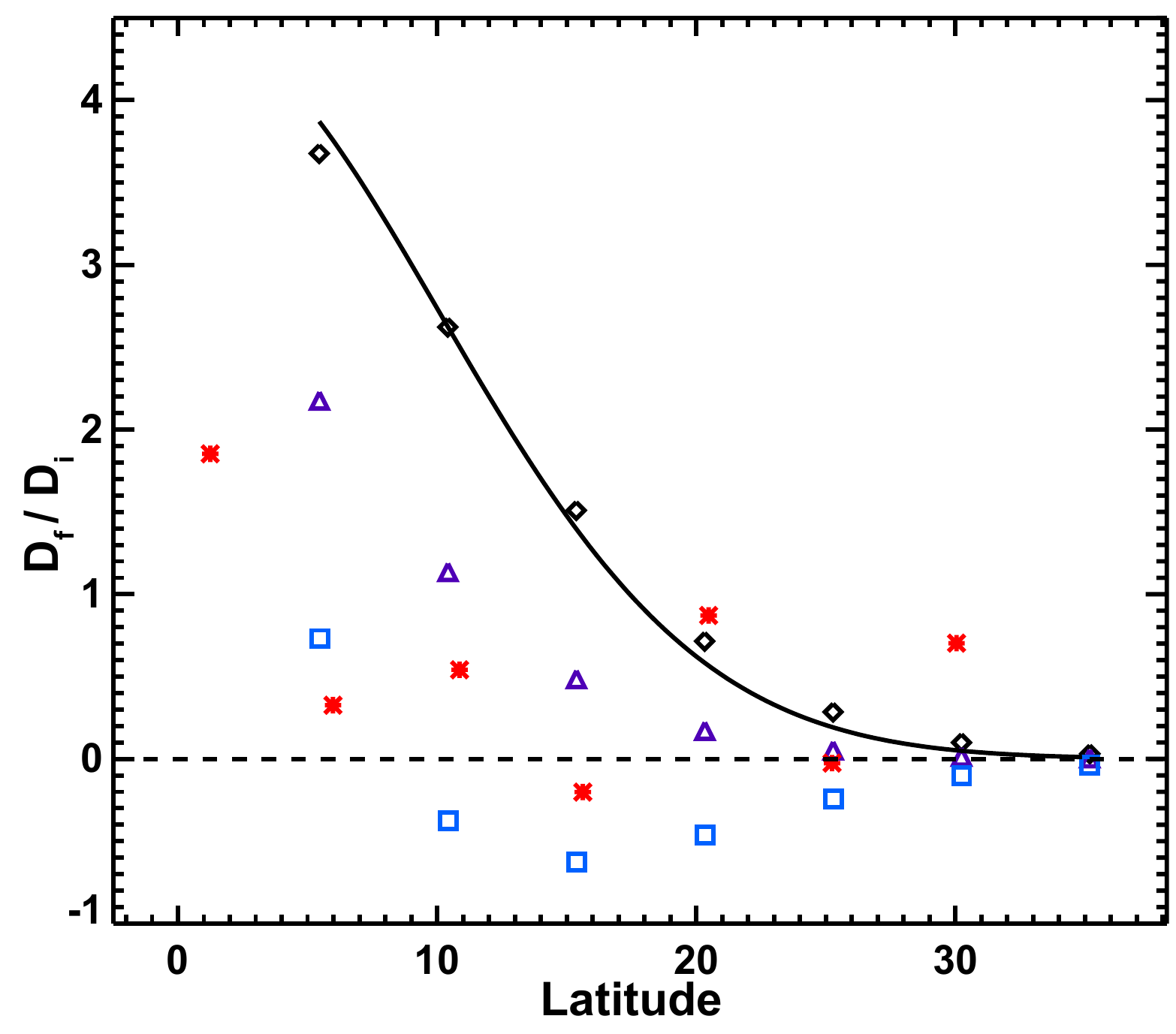}
\caption{Final dipole contribution factor $D_{f}/D_{i}$.  The black diamonds represent the results of the symmetric bipolar ARs. The purple triangles represent the results of the weakly asymmetric bipolar ARs.  The blue squares represent the results of the strongly asymmetric bipolar ARs.  The red asterisks represent the results of the complex ARs.  The black trend line representing Eq. (\ref{bmr1}) is fitted to the symmetric bipolar ARs.  The black dashed horizontal line marks $D_{f}/D_{i}=0$.\label{fig:fvaluebmr}}
\end{figure*}

The artificially created ARs are simulated to get their corresponding final dipole contributions.  We also calculate the results of the two algebraic prediction methods for the ARs as described in Sect. \ref{subsec:evaars}.  The corresponding contribution factors $D_{f}/D_{i}$ are plotted in Fig. \ref{fig:fvaluebmr}.  As the bipolar ARs at the equator do not have an initial dipole because of the zero tilt angle, they are not shown here.  As shown, the different cases do not follow a uniform trend.  The bipolar ARs with different degrees of asymmetry (marked with black diamonds, purple triangles, and blue squares) appear to show separate trends.  The bipolar ARs with more diffusive following polarities are systematically overestimated by the BMR-based method, as they all appear under the trend line.  This indicates that they contribute less to the final dipole moment as more flux of the following polarity tends to migrate across the equator, which agrees with the results of \citet{2019ApJ...883...24I}.  For the strongly asymmetric cases, the contributions predicted by the BMR-method are even opposite to those suggested by SFT simulations at some latitudes.  The results of the complex ARs based on the configuration of AR 12673 (marked with red asterisks) do not follow a trend at all, and the contribution factor even becomes negative for some cases, as first shown by \citet{2019ApJ...871...16J}.  In summary, for the asymmetric bipolar ARs illustrated by the artificially created ones, the BMR-based method systematically overestimates the dipole contribution, while for the complex ARs, the dipole contribution given by the BMR-based method does not show a systematic trend.

The results of the artificially created ARs are consistent with the results of observed ARs in Sect. \ref{subsec:evaars}.  From the perspective of Eq. (\ref{math5}), the major contribution of an AR to the final dipole originates from its components with lower latitudes and higher fluxes, instead of the properties of the whole AR.  Hence, for ARs with irregularly distributed flux, the final dipole contributions deviate from the predictions of the BMR-based method, which  only considers the overall properties of the ARs.  The deviation depends on the exact configuration of the ARs instead of a systematic property.  Even for ARs with the same shape, a slight change in latitude causes a substantial alteration of the contributions, as shown by the red asterisks in Fig. (\ref{fig:fvaluebmr}).  As many observed ARs are not bipolar in reality, the BMR-based method, instead of revealing a systematic deviation, tends to randomly miscalculate the results.

\begin{figure*}
\centering
\includegraphics[width=0.7\textwidth]{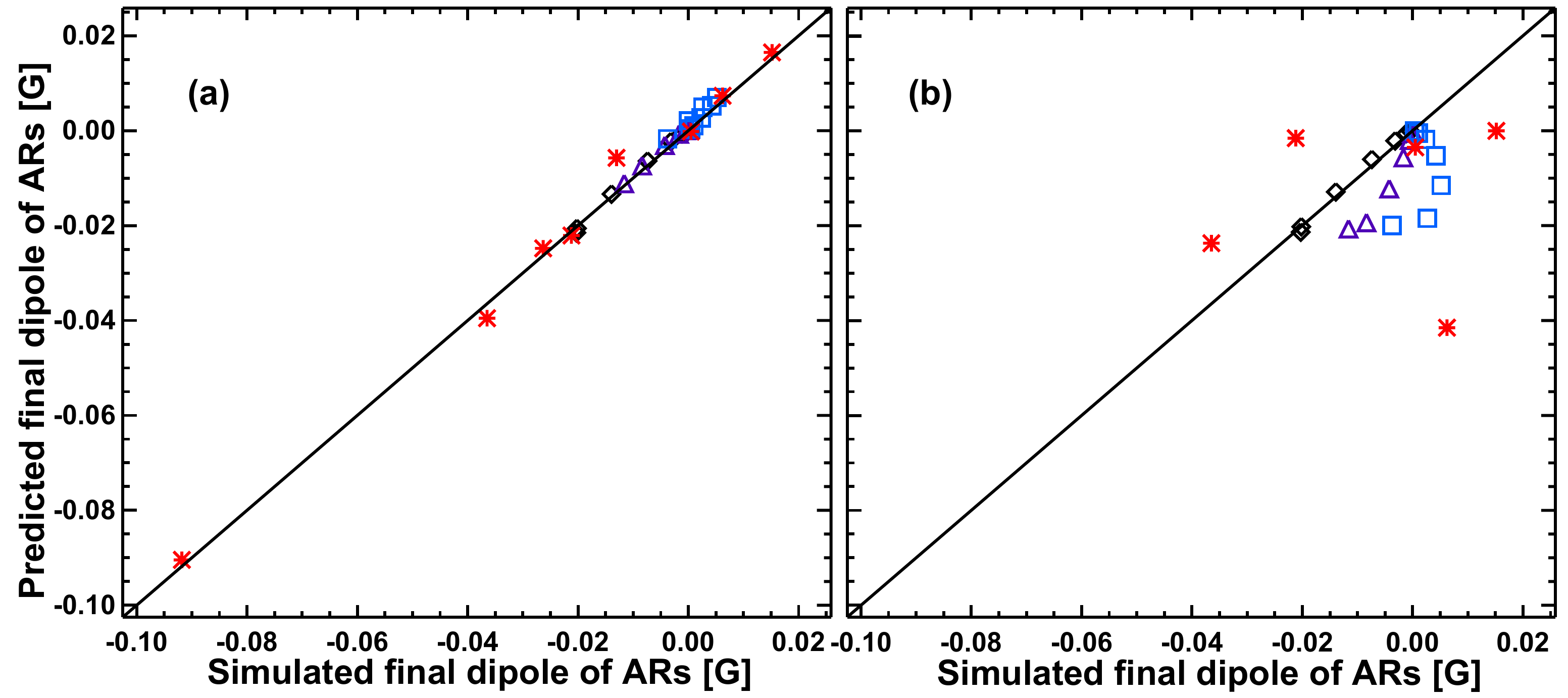}
\caption{Same as Figure \ref{fig:dipar}, but for artificially created ARs.  The black diamonds represent the results of the symmetric bipolar ARs. The purple triangles represent the results of the weakly asymmetric bipolar ARs.  The blue squares represent the results of the strongly asymmetric bipolar ARs.  The red asterisks represent the results of the complex ARs.  Panel (a) presents the results of the generalized method that we introduce; panel (b) presents the results of the BMR-based method.  We note that for panel (b), three points of the complex ARs lie beyond the display range, with the most extreme point (-0.092,-0.22).\label{fig:dipbmr}}
\end{figure*}

Now we deal with the four kinds of configurations of ARs located at different latitudes in the same way as Fig. \ref{fig:dipar}.  The results are shown in Fig. \ref{fig:dipbmr}.  Figure \ref{fig:dipbmr}(a) indicates that the generalized algebraic method still produces reliable final dipole predictions for all cases.  The different configurations of the ARs do not  notably affect the results.  On the contrary, the BMR-based method cannot produce reliable predictions for the asymmetric or complex cases (Fig. \ref{fig:dipbmr}(b)).  The asymmetric bipolar ARs (purple triangles and blue squares) deviate from the main trend of the symmetric BMRs, and the complex ARs (red asterisks) are completely irregularly distributed.  As clarified above, this can be explained by the fact that the parameter $\lambda_{R}$ is not fixed when the initial configuration of the flux patches is not as diffusive as that of other ARs, and is completely ill-defined for complex ARs.  The generalized method on the other hand  is not affected by this issue, as it does not involve BMR assumptions.  The generalized method maintains its accuracy regardless of different AR configurations.  We explicitly show that the generalized method is a more generalized approximation of the SFT model, while the previous BMR-based method considers only a special case.



\section{Discussion and conclusions} \label{sec:outro}


We introduce a generalized algebraic method to calculate the contribution of an arbitrary AR to the axial dipole moment at cycle minimum, which is the precursor of the strength of the next cycle.  The method is a time-efficient alternative to the SFT simulations for prediction of the final dipole contributions of the ARs.  The method describes the cross-equatorial flux transport originating from a point magnetic flux patch, i.e., described by a Dirac $\delta$ function profile flux, and deduces the associated final dipole moment contribution.  The contribution of an arbitrary AR to the final dipole moment is an integration of the contributions of the  points covering the whole AR area.  This method is a generalization of the previous Gaussian contribution profile method, which assumes a flux emergence in the form of BMRs with fixed width \citep{2020JSWSC..10...50P}.  By comparing to simulations of artificially created
BMRs with different degrees of asymmetry and more complex ARs, we explicitly show that the generalized method is not affected by AR configurations as the BMR-based method is.  The generalized method produces more accurate dipole predictions compared to the BMR-based method for realistic ARs, and requires considerably less computation time than SFT simulations.  On the contrary, the BMR-based method tends to systematically overestimate the contributions of asymmetric bipolar ARs, and randomly miscalculate the contributions of ARs with complex configurations.  Hence, the generalized method offers a quick alternative to SFT simulations for precisely quantifying the  influence of an AR on solar cycle evolution, and can be used to accelerate and improve physics-based solar cycle predictions.

While the generalized algebraic method we introduce can precisely predict AR contributions to the final dipole moment, the algebraic method does not produce the whole process of the development of the large-scale magnetic field, and so SFT simulations are still required to understand the development of solar magnetism at the solar surface and the heliosphere.  The algebraic method does not replace realistic SFT simulations.

The generalized algebraic method is valid for any linear kinematic SFT model regardless of transport parameters and profiles.  An SFT model with different assumptions of the transport terms can also be described by the algebraic method, but the exact parameters of the method may be different.  Therefore, the method is universal for SFT models, but the results of the method are limited to a specific SFT model with a set of given transport parameters.  When we use the algebraic method to quantify AR contributions to the solar cycle in the future, the parameters $\lambda_{R}$ and $A_{0}$ should be derived from the transport parameters of the specific SFT model we adopt.  

The generalized algebraic method is a theoretical approximation and a quick implementation of the SFT model, while the previous BMR-based method is its special case.  The SFT model describes the emerged ARs after reaching their decay phase, not the flux emergence of ARs.  The generalized algebraic method cannot describe the emergence of ARs either.  Therefore, the ARs should be accurately identified on their decay phase to precisely quantify their contributions to the solar cycle.  For ARs that emerge in spatial and temporal proximity to others, automatically identifying their exact decay phase can be problematic, as the fluxes of different ARs often superpose one another as they evolve \citep{2020ApJ...904...62W}.  This problem should be considered in the analysis and characterization of the influence of ARs on the solar cycle.


The generalized method requires the real configurations of ARs, and so the spatial resolution of the input ARs needs to be carefully considered. To what extent the details of the ARs is required is unclear.  Theoretically, the SFT model implements the supergranular diffusion, which is introduced by the random walk process from the supergranulation convective currents \citep{1964ApJ...140.1547L}.  This potentially limits the resolution of the SFT model and its algebraic equivalent method.  Further numerical and observational studies are needed to clarify the required  resolution of ARs, such as the previous efforts of \citet{2001ApJ...547..475S}, \citet{2016ApJ...830..160M}, and \citet{2018GeoRL..45.8091U}.  These studies may reveal the effective range of the diffusion approximation for the magnetic flux evolution at the solar surface.

\begin{acknowledgements}

We thank the referee for the valuable comments and suggestions on improving the manuscript.  The SDO/HMI data are courtesy of NASA and the SDO/HMI team.  This research was supported by the Key Research Program of Frontier Sciences of CAS through grant No. ZDBS-LY-SLH013, National Natural Science Foundation of China through grant Nos. 11873023, 11322329, and 11533008, and the B-type Strategic Priority Program of CAS through grant No. XDB41000000.  J.J. acknowledges the International Space Science Institute Teams 474 and 475.

\end{acknowledgements}

\end{document}